# SOLVING FUNCTIONAL RELIABILITY ISSUE FOR AN OPTICAL ELECTROSTATIC SWITCH

*H. Camon[1], C. Ganibal[1], N. Rapahoz[2], M. Trzmiel[2], C. Pisella[2], C. Martinez[3], K. Gilbert[3], S. Valette[3]*

[1] LAAS/CNRS – 7 avenue du colonel Roche, 31077 Toulouse cedex 4, France
[2] Tronic's Microsystems – 55 rue du Pré de l'Horme, 38926 Crolles cedex, France
[3] CEA-LETI/DOPT – 17 rue de Martyrs, 38054 Grenoble cedex 9, France

**ABSTRACT**

In this paper, we report the advantage of using AC actuating signal for driving MEMS actuators instead of DC voltages. The study is based upon micro mirror devices used in digital mode for optical switching operation. When the pull-in effect is used, charge injection occurs when the micro mirror is maintained in the deflected position. To avoid this effect, a geometrical solution is to realize grounded landing electrodes which are electro-statically separated from the control electrodes. Another solution is the use of AC signal which eliminates charge injection particularly if a bipolar signal is used. Long term experiments have demonstrated the reliability of such a signal command to avoid injection of electric charges.

## 1. INTRODUCTION

MEMS (Micro Electro Mechanical Systems) and MOEMS (Micro Opto Electro Mechanical Systems) devices are yet widely broadcasting even towards common industrial applications. One of the most spectacular examples is the micro-mirror matrix for video displays (DMD from Texas Instruments). In telecommunications, the use of the MEMS for the optical switching like Optical-Cross-Connect or wavelength manipulation like Add-and-Drop has generated a lot of work since the end of the 90s with a plentiful literature showing spectacular results in term of performance [1, 2, 3, 4, 5 and 6]. Within the framework of the project ROADMAP [7], we have developed a wide family of switch on base of digital mirrors that must simplify the electronics of command and allow a big inconstancy of conception.

The figure 1 gives the principal plan. The first digital mirror of chip A allows to select one of the two input fibers (function 2X1) while the first mirror of the chips B selects one of the two output fibers (function 1x2). Intermediate optics insure the management of the sizes of optical mode in free space propagation (Gaussian conjugation), and the realignment of deviated beams by mirrors (geometrical conjugation). To clarify the choice of digital mirrors, we are going to present shortly the behaviour of electrostatic micro mirrors illustrated by figure 2 [8, 9 and 10].

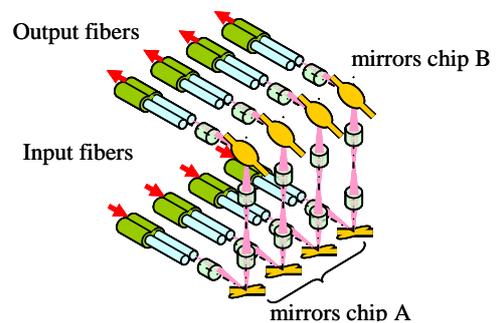

Figure 1. Schematic of an optical switch

A micro mirror is constituted by a rigid reflective plate suspended by two arms of torsion above two electrodes of command (a right and a left). The application of a voltage between the micro mirror generally grounded and one of the electrodes of command generates a electrostatic force bringing the micro mirror to rotate in the chosen direction. The angular position of the mirror is determined by an equilibrium state between electrostatic forces and restoring mechanical torque. When a small voltage is applied, the angle of rotation, α, varies linearly with the squared value of the applied voltage. By increasing the amplitude of the input voltage, the response of the system becomes non linear. From a specific voltage, Vpi, and above the electrostatic moment becomes always more important than the mechanical restoring torque and the micro mirror rotates to the maximum angle geometrically allowed by the structure, $\alpha_{max}$, and the mirror is then in contact with the underlying layer. This is the pull-in effect determined by the pull-in voltage (Vpi) used in the digital mode. This means that the mirror could be rotate, when it is statically driven, only to corresponding angle of Vpi, the pull-in angle $\alpha_{pin}$. A small increase in input voltage results in a direct rotation of the mirror to the angle $\alpha_{max}$. We can observe a forbidden region in angle domain between $\alpha_{pin}$ and $\alpha_{max}$. At this point, when the voltage is progressively reduced, the deflection angle stays at this maximum value, $\alpha_{max}$, until the mechanical restoring torque overcomes the electrostatic one at the maintaining voltage (Vm). For lower voltages, the angle follows again a linear relationship with the applied voltage.

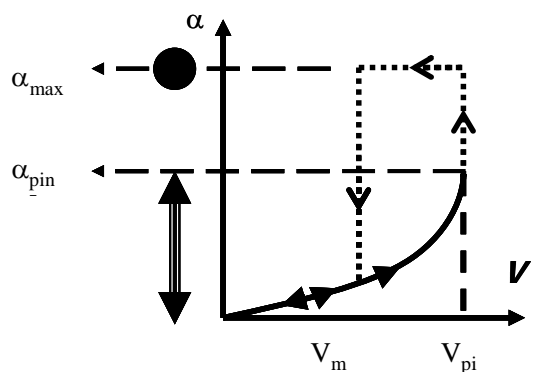

Figure 2. Static angular behaviour versus the amplitude of voltage applied

Micro mirror behaviour could be divided in two regions. In the first one, called analogical mode region, the voltage applied is lower than Vpi. The micro mirrors can be actively positioned (with each deflection angle corresponding to an applied voltage) until the operating frequency of the input is lower than the mechanical resonance frequency. In other case, the micro mirror can not follow the input signal. In the second one, called digital region, any voltage applied results in only one deflection angle, $\alpha_{max}$. The conception of the system wants to take advantage of the pull-in effect. In the analogical region, all the angles between 0° and $\alpha_{pin}$ are obtainable but at the price of the preservation of the balance of attraction and back moving forces. It requires setting up a fast control loop to guarantee the maintenance of the angle. In addition, the relation between angle and applied voltage is not linear. This makes more complex the definition of the law of command. In an opposite way, in the digital region, a single voltage of actuation is required but for a single angle. There is thus no more a control of position. Another benefit is the use of the maximum angle allowed by the structure ($[0, \alpha_{pin}] < \alpha_{max}$). Also, a large angle could always be seen as an addition of constant small angles. In other words, the large number of angular position authorized by the analogical systems can be implemented by the assembly of a series of smaller constant angles. The price is an increase of the mirror number and the increasing requirements for assembly.

## 2. STRUCTURE OF THE MIRROR AND DYNAMIC BEHAVIOUR

The figure 3 shows the cross-section of mirrors MEMS which was developed by the partner company (Tronic's Microsystems) of the project. It is constituted by 4 mirrors of 600x600 $\mu m^2$ in size suspended by torsion beams arms above electrodes for command. The figure 3 gives the general principle of manufacturing. The lower silicon substrate is fabricated by wet etching to obtain a balance-knife-edge shape with a bottom located at h µm from the initial surface. Then the command electrodes are deposited and electrically isolated by a low thermal silicon oxide layer. The both faces of the SOI upper wafer are also micro machined using wet etching in order to obtain the mirror and the two torsion beams with the same thickness, 10 µm in our case. The mirror face is covered by a gold layer to enhance reflectivity. The both wafer are then aligned, glued together and finally sawed in individual pieces with four micro mirrors. The figure 4 illustrates the behaviour of such designed mirror. When increasing actuation voltage, the mirror slightly rotate to a value of 0.5 degree (analogical region) and at Vpi reach the value of 1.6 degree corresponding to $\alpha_{max}$, the final position used in digital mode. When the applied voltage is decreased, $\alpha_{max}$ is maintained until $V_m$. The figure 5 illustrates an example of commutation between two adjacent output fibres in a complete system.

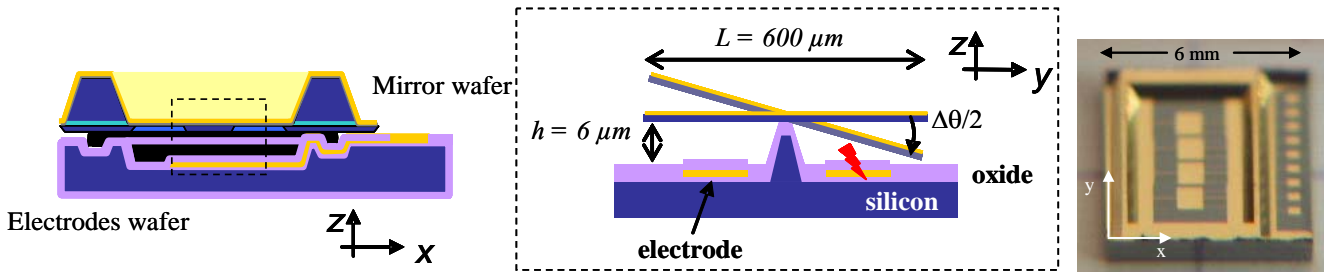

Figure 3. Structure of Tronic's chips:
to the left, cross-section along the axis of rotation of mirrors; in the middle: partial cross-section in a perpendicular plane of the rotation axis of mirrors; to the right: photograph of a set of four mirrors (5x5 mm²)

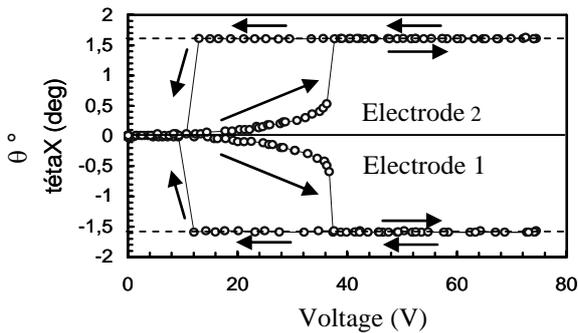

Figure 4. right and left angle versus voltage for one mirror

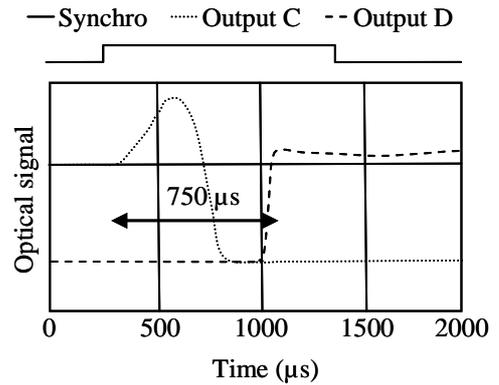

Figure 5. Commutation from one fibre to an adjacent one

## 3. MECHANICAL AND FUNCTIONAL RELIABILITY

Ideally, optical switch could be used at different level of the communication network but the features of functioning vary according to this level. At the backbone level, the number of switching operation is weak (some per year) while, the more we get closer to the user, the more the number of switching operation increases. Whatever the level, the switching time always remains equal to one millisecond and the life time of components oblige to have an excellent mechanical reliability. We can easily verify this point. For mechanical reliability, long term experiment has been carried on. A micro mirror is tilted from one side to the other during more than 460 days with a period of 2 ms corresponding to 3.97 $10^{10}$ switching operation without any modification of switching time (Figure 6). This result confirms the excellent mechanical reliability of mono crystalline silicon. This point is mandatory because a system must be guaranteed to work several years without any modifications of behaviour.

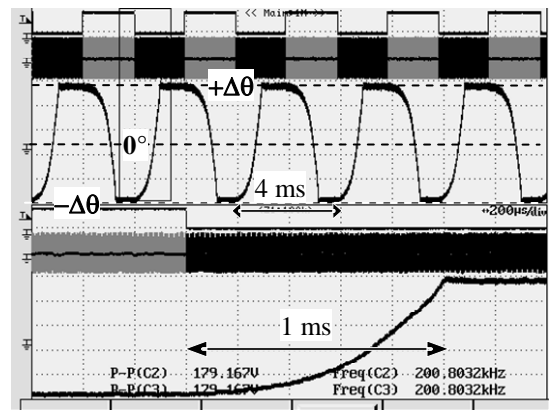

Figure 6. Long term experiment for testing mechanical reliability: in the upper part from the top to the bottom: TTL signal, applied command signals to electrodes in black and gray, signal from the PSD; lower part: zoom from the upper part)

For the functional reliability, we have to verify that keeping the mirror in tilted position for a long time does not affect the next switching. This is the case for operation at backbone level. Figure 7 illustrates experiments on a micro-mirror device. A laser beam is focused on the micro-mirror and after reflection intercepts a positioning sensor device (PSD). We apply a triangular DC voltage signal trigged by a TTL signal. The frequency is 4 Hz and amplitude 0-100 V. The values of $V_{pi}$ and $V_m$ can be obtained during the increasing half period and the decreasing half period respectively by detecting rotation to $\alpha_{max}$ (detection of $V_{pi}$) and return to 0° (detection of $V_m$).

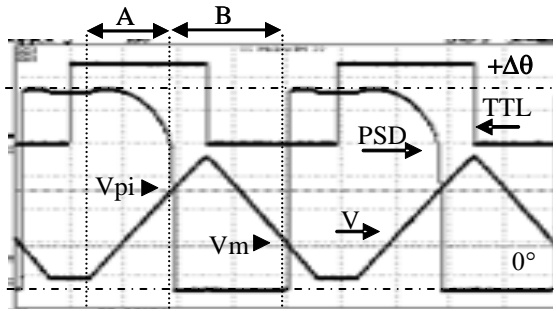

Phase A: Static deviation
Phase B: commutation ($V_{pi}$) and release ($V_m$)

Figure 7. Angular position of mirror actuated with a slow triangular input voltage

Figure 8 illustrates the evolution of $V_{pi}$ and $V_m$ during a long term experiment on a micro mirror device. It is clear that tilting the mirror becomes more and more difficult as the number of actuations increases. The voltage needed to swing the mirror increases by 26 V with time. The same value (exactly 23 V) of increase is observed for Vm with the same evolution.

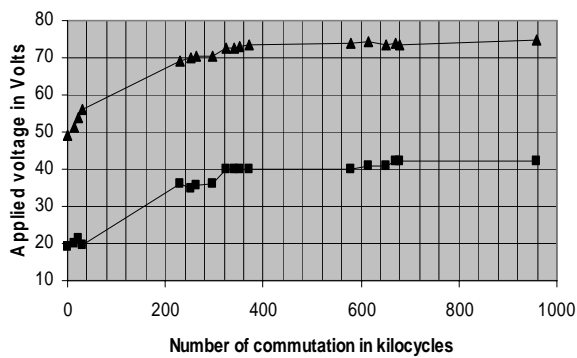

Figure 8. Evolution of Vpi (triangle dots) and Vm (square dots) in time

This increase in voltage is due to the injection of electrons in the oxide layer. This effect can be accelerated by applying square DC signal resulting, at the end, in blocking the mirror in tilted position even when all electrodes are grounded [11]. In tilted position, the mirror touches the oxide layer. In this position, the electric field between the mirror and the electrode is at its maximum (roughly 1 MV/cm) and injects electric charges in it. Then, when electrodes are grounded, the injected electric charge is strong enough to maintain the mirror in the tilted position. This is in agreement with electromechanical simulation and capacitance versus voltage experiment.

To by-pass this difficulty, we used bipolar signals with high frequency. A bipolar signal to withdraw injected electric charges by reversal of the electric field during the half time of application, and, at enough high frequency (above the mechanical resonance of the mirror) to avoid movement of the mirror. The mechanical structure of the mirror acts like a low-pass filter. Figure 9 illustrates the efficiency of the method. An unused mirror is tilted one time and the return to 0° position is snapshot. Then a continuous high frequency voltage (70 kHz) is applied to maintain the mirror in the tilted position. From time to time (every hours at the beginning and every day later), the electrodes are grounded for 2 ms allowing to snapshot the return to 0° position. The command signal is immediately applied after each 2 ms interruption.

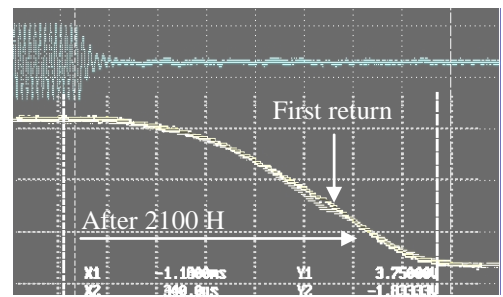

Figure 9. Overlap of snapshoot return-to-0° trajectory (bipolar HF command signal in blue, snapshots in white)

During the return-to-0° phase, electrodes are grounded and the dynamic is only govern by mechanical restoring torque and electrical forces induced by injected charges. The absence of difference between the initial trajectory and the following ones indicates clearly the absence of electric charges or at least to a very weak level not perturbing the nominal functioning of the mirror.

## 4. CONCLUSION

On the basis of experimental results we can conclude that using bipolar AC signal presents advantages for devices using MEMS digital micro-mirrors. In digital mode and in the tilted position, that is when the mirror touches the electrode level, the electric field in the insulating layer between the mirror and the electrodes is maximised. The consequence is the progressive injection of electric charges leading to the blocking of the mirror in tilted position. Thus the use of alternating bipolar signal permits to swap the electric field and to minimise, even to cancel, electric charge injection. This minimisation allows maintaining the mirror in the tilted position for a long time and insuring the return to the neutral position when the voltage is deactivated. Actually the functional reliability of the system is sharply improved compared with a command by DC signal.

In addition, the use of bipolar AC signals permits to simplify the design and the fabrication. It is not necessary to design grounded landing electrodes minimising connections. This is applicable to all MEMS where contact between the moving part and an electrode exists. This point was tested during experimentation over several months showing the efficiency of the approach. The resulting injected charges can be considered negligible. This method allows to release an operational optical cross connect.

## 5. REFERENCES

[1] V.A. Aksyuk et al., J. Lightwave Technol. Vol 21, n°3, 2003, pp. 634-642.

[2] L-S. Fan, Y.C. Tai, R.S. Müller, "IC-processed electrostatic micromotors", Sensors & Actuators 20, 1989, pp. 41-47.

[3] J.M. Busillo, R.T. Howe and R.S. Muller, "Surface micromachined for Micro electro mechanical Systems", Proceedings of the IEEE 86, 1998, pp. 1552-1574.

[4] D. W. Monk and R. O. Gale, "The digital micromirror device for projection display", Microelectronic Engineering 27, 1995, pp. 489-493.

[5] W. M. Duncan, T. Bartlett, B. Lee, D. Powell, P. Rancuret and B. Sawyers, "Dynamic optical filtering in DWDM systems using the DMD", Solid-State Electronics 46, 2002, pp. 1583-1585.

[6] C. Lee, "Monolithic-integrated 8CH MEMS variable optical attenuators", Sensors and Actuators A 123-124, 2005, pp. 596-601.

[7] C.Martinez, K.Gilbert, S.Valette, N.Raphoz, M.Trzmiel, C.Pisella, H.Camon, C.Ganibal, T.Zami and M.Prunaire, "Réalisation d'un commutateur optique espace libre à base de MEMS digitaux", JNOG'2006, Metz, 2006, pp.418-420.

[8] X. M. Zhang, A. Q. Liu, C. Lu, F. Wang and Z. S. Liu, "Polysilicon micromachined fiber-optical attenuator for DWDM applications", Sensors and Actuators A 108, 2003, pp. 28-35.

[9] H. Camon and F. Larnaudie, "Fabrication, simulation and experiment of a rotating electrostatic silicon mirror with large angular deflection", $13^{th}$ Int.. Micro Electro Mechanical Systems (MEMS), Miyazaki, 2000.

[10] H. Camon, F. Larnaudie, C. Ganibal and B. Estibals, "Scanning system using an 1D silicon micro-mirror", 3rd Topical Meeting on Optoelectronic Distance Measurement and Applications, Pavia, September 20-22, pp. 2001.

[11] Z. Liu, M. Kim, N. Y-M. Shen and E. C. Kan, "Actuation by electrostatic repulsion by nonvolatile charge injection", Sensors and Actuators A 119, 2005, pp. 236-244.